\documentclass[aps,preprint,showpacs,preprintnumbers,eqsecnum,amsmath,amssymb]{revtex4}

\usepackage{graphics,epsfig}
\usepackage{graphicx}
\usepackage{dcolumn}
\usepackage{bm}

\begin{document}

\title{Late-time evolution of charged massive Dirac fields in
the Reissner-Nordstr\"om black-hole background}
\author{Jiliang Jing} \email{jljing@hunnu.edu.cn}
\affiliation{ Institute of Physics and  Department of Physics, \\
Hunan Normal University,\\ Changsha, Hunan 410081, P. R. China }

\begin{abstract}

The late-time evolution of the charged massive Dirac fields in the
background of a Reissner-Norstr\"om (RN) black hole is studied. It
is found that the intermediate late-time behavior is dominated by
an inverse power-law decaying tail without any oscillation in
which the dumping exponent depends not only on the multiple number
of the wave mode but also on the field parameters. It is also
found that, at very late times, the oscillatory tail has the decay
rate of $t^{-5/6}$ and the oscillation of the tail has the period
$2\pi/\mu$ which is modulated by two types of long-term phase
shifts.

\end{abstract}


\maketitle

\section{introduction}

The evolution of field perturbation around a black hole consists
roughly of three stages \cite{Frolov98}. The first one is an
initial wave burst coming directly from the source of
perturbation. The second one involves the damped oscillations
called the quasinormal modes. And the last one is a power-law tail
behavior of the waves at very late time.

The late-time evolution of various field perturbations outside a
black hole has important implications for two major aspects of
black-hole physics: the no-hair theorem and the mass-inflation
scenario\cite{Poisson}\cite{Burko}. Therefore, the decay rate of
the various fields has been extensively studied
\cite{3}-\cite{Ching} since Wheeler \cite{1,2} introduced the
no-hair theorem. Price \cite{3} studied the massless external
perturbations and found that the late-time behavior for a fixed
~$r$ is dominated by the factor $t^{-(2l+3)}$. Barack, Ori and Hod
considered \cite{7}-\cite{9} the late-time tail for the
gravitational, electromagnetic, neutrino and scalar fields in the
Kerr spacetime. Starobinskii and Novikov \cite{Starobinskii}
analyzed the evolution of a massive scalar field in the RN
background and found that there are poles in the complex plane
which are closer to the real axis than in the massless case. Hod
and Piran \cite{13} pointed out that, if the field mass $\mu$ is
small, the oscillatory inverse power-law behavior $ \Phi\sim
t^{-(l+3/2)}\sin(\mu t)$ dominates as the intermediate late-time
tails in the RN background. We \cite{Jing1} studied the late-time
tail behavior of massive Dirac fields in the Schwarzschild
black-hole geometry and found that this asymptotic behavior is
dominated by a decaying tail without any oscillation. Koyama and
Tomimatsu \cite{14} found that the very late-time tail of the
massive scalar field in the Schwarzschild and RN background is
approximately given by ~$t^{-5/6}\sin(\mu t)$.

Although much attention has been paid to the investigations of the
late-time behaviors of the neutral scalar, gravitational and
electromagnetic fields in the static and stationary spacetimes, at
the moment the study of the late-time tail evolution of the
charged massive fields is still an open question. The main purpose
of this paper is to study the late-time tail evolution of the
charged massive Dirac fields in the RN black-hole background.

\section{Late-time tail of the charged massive Dirac fields}

In the RN spacetime the Dirac equations coupled to a
electromagnetic fields \cite{Page} can be separated by using the
Newman-Penrose formalism \cite{Newman}. After the tedious
calculation, we find that the angular equation is the same as in
the Schwarzschild black hole \cite{Jing1} and the radial equation
can be expressed as
\begin{eqnarray}
&&  \frac{d^2 \Psi_{\pm}}{d r_*^2}+\left\{\frac{d H_{\pm}}{d
r_*}-H_{\pm}^2+\frac{\Delta}{r^4}P_{\pm}\right\}\Psi_{\pm}
 =0,\label{LV1}
 \end{eqnarray}
with
 \begin{eqnarray}
H_{\pm}&=&\mp \frac{1}{4
r^2}\frac{d\Delta}{dr}-\frac{\Delta}{r^3}\mp
\frac{i\mu}{2(\lambda\mp i\mu r)}\frac{\Delta}{r^2}, \nonumber \\
 P_{\pm}&=&\frac{K^2-isK\frac{\Delta}{dr}}{ \Delta} +4 i s \omega
r-2 i s e Q+2 s(s+\frac{1}{2})
\nonumber \\
&& +\frac{\mu[\frac{i}{2}(s+\frac{1}{2})\frac{d
\Delta}{dr}-K]}{\lambda\mp i\mu r}
 -\mu^2r^2 -\lambda^2,\nonumber
 \end{eqnarray}
where $\Delta=r^2-2M r +Q^2$ ($M$ and $Q$ represent the mass and
charge of the black hole), $ dr_*=(r^2/\Delta) dr$ and $\Psi_{\pm
}=r\Delta^{\pm \frac{1}{4}}(\lambda^2+\mu^2
r^2)^{-\frac{1}{4}}e^{\pm  i \arctan(\frac{\mu r}{\lambda})/2}
R_{\pm \frac{1}{2}}$  ($R_{\pm \frac{1}{2}} $ is an usual radial
wave function \cite{Jing1}).

Let us assume that both the observer and the initial data are
situated far away from the black hole. Then, we can expand Eq.
(\ref{LV1}) as a power series in $M/r$ and $Q/r$ and obtain
(neglecting terms of order $O\left(\left(\omega/r\right)^2\right)$
and higher)
\begin{eqnarray}\label{Pq}
\label{19}&&\left[\frac{d^{2}}{dr^{2}}-\varpi^{2}+ \frac{2 a
\varpi }{r} -\frac{b^2-\frac{1}{4} }{r^{2}} \right] \xi_{\pm}=0,
\end{eqnarray}
where $\varpi=\sqrt{\mu^2-\omega^2}$, $a=(M\mu^2+e
Q\omega)/\varpi-2M\varpi $,
$b^2=\frac{1}{4}+\lambda^2+4M\mu^2-Q^2(e^2+\mu^2)-2is e
Q+(\lambda/\mu+8MeQ+2is M)\omega$ and
$\xi_{\pm}=(\Delta/r^2)^{1/2}\Psi_{\pm}$. We obtain the two basic
solutions required to build Green's function
\begin{eqnarray}
\label{mgreen} &&\tilde{\Psi}_1=Ae^{-\varpi r}(2\varpi r)
^{\frac{1}{2}+b}M(\frac{1}{2}+b-a, 1+2 b, 2\varpi r),\nonumber \\
&&\tilde{\Psi}_2=Be^{-\varpi r}(2\varpi
r)^{\frac{1}{2}+b}U(\frac{1}{2}+b-a, 1+2 b, 2\varpi r),
\end{eqnarray}
where $A$ and $B$ are normalization constants,
$M(\tilde{a},\tilde{b},z)$ and $U(\tilde{a},\tilde{b},z)$
represent the two standard solutions to the confluent
hypergeometric equation \cite{Abramowitz}.
$U(\tilde{a},\tilde{b},z)$ is a many-valued function, i.e., there
is a cut in $\tilde{\Psi}_2$. Hod, Piran and Leaver \cite{13, 17}
found that the asymptotic massive tail is associated with the
existence of a branch cut (in $\tilde{\Psi}_{2}$) placed along the
interval $-\mu\leq \omega\leq\mu$ and the branch cut contribution
to the Green's function is
\begin{eqnarray}
\label{mgreen1} G^C(r_*,r_*';t) =\frac{1}{2\pi}\int_{-\mu}^\mu
F(\varpi) e^{-i\omega t}d \omega.
\end{eqnarray}
with

\begin{eqnarray}
&&F(\varpi)=\nonumber \\ &&\frac{\tilde{\Psi}_1(r_*',\varpi
e^{i\pi})\tilde{\Psi}_2(r_*,\varpi e^{i\pi})} {W(\varpi e^{i
\pi})}-\frac{\tilde{\Psi}_1(r_*',\varpi)
\tilde{\Psi}_2(r_*,\varpi)} {W(\varpi)},\nonumber
\end{eqnarray}
where  $W(\varpi)=W(\tilde{\Psi}_{ 1},\tilde{\Psi}_{
2})=\tilde{\Psi}_{ 1}\tilde{\Psi}_{ 2,x}-\tilde{\Psi}_{ 2}
\tilde{\Psi}_{1,x}$ is the Wronskian.  We obtain, with the help of
Eq. (13.1.22) of Ref. \cite{Abramowitz}, that $W(\varpi
e^{i\pi})=-W(\varpi)=AB\frac{\Gamma(2b)}{\Gamma(1/2+b-a)} 4
b\varpi$.
 For simplicity we assume that the initial data has a
considerable support only for $r$ values which are smaller than
the observer's location. This, of course, does not change the
late-time behavior. Noting that when $t$ is large, the term
$e^{-i\omega t}$ oscillates rapidly. This leads to a mutual
cancellation between the positive and the negative parts of the
integrand, so that the effective contribution to the integral
arises from $|\omega|=O(\mu-\frac{1}{t})$ or equivalently
$\varpi=O(\sqrt{\mu/t})$ \cite{13}. Using Eqs. (13.1.32),
(13.1.33) and (13.1.34) of Ref. \cite{Abramowitz}, we find that
$F(\varpi)$ is given by

\footnotesize
\begin{widetext}
\begin{eqnarray}\label{FF}
\label{F1}F(\varpi)&=&\frac{ r_*^{\frac{1}{2}-b}
r_*'^{\frac{1}{2}+b}e^{-\varpi (r_*+r_*')}}{2
b}[M(\frac{1}{2}+b+a,1+2b,2\varpi r_*')M(\frac{1}{2}-b+a,1-2b,
2\varpi r_*) -M(\frac{1}{2}+b-a,1+2b,2\varpi r_*') \nonumber
\\ &&  \times M(\frac{1}{2}-b-a,1-2b,2\varpi
r_*)] -\frac{\Gamma(-2b)\Gamma(\frac{1}{2}+b-a)}{\Gamma(2b)
\Gamma(\frac{1}{2}-b-a)}\frac{(4\varpi^2 r_* r_*')
^{\frac{1}{2}+b}e^{-\varpi (r_*+r_*')}}{4\varpi
b}[M(\frac{1}{2}+b-a,1+2b,2\varpi r_*')\nonumber
\\ && \times M(\frac{1}{2}+b-a,1+2b,
2\varpi r_*)  +   e^{(1+2b)i\pi}M(\frac{1}{2}+b+a,1+2b,2\varpi
r_*')M(\frac{1}{2}+b+a,1+2b, 2\varpi r_*) ].
\end{eqnarray}
 \end{widetext}
\normalsize

\begin{figure}
\includegraphics[scale=0.8]{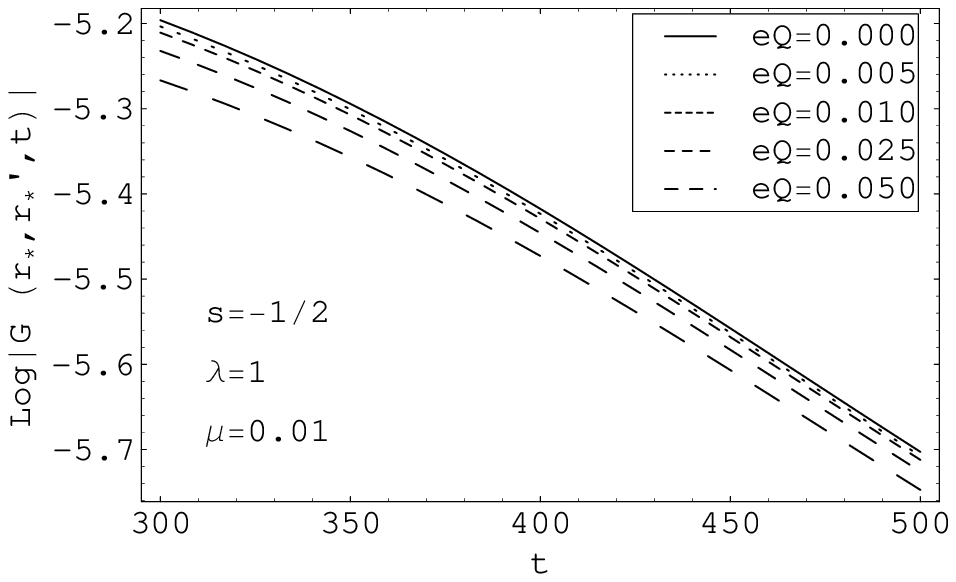}
\includegraphics[scale=0.8]{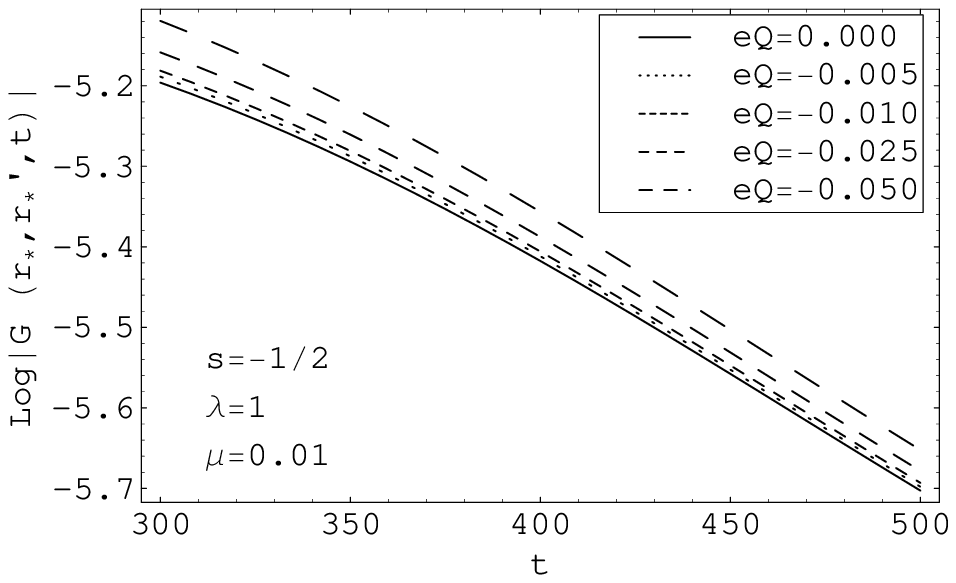}\\
\includegraphics[scale=0.8]{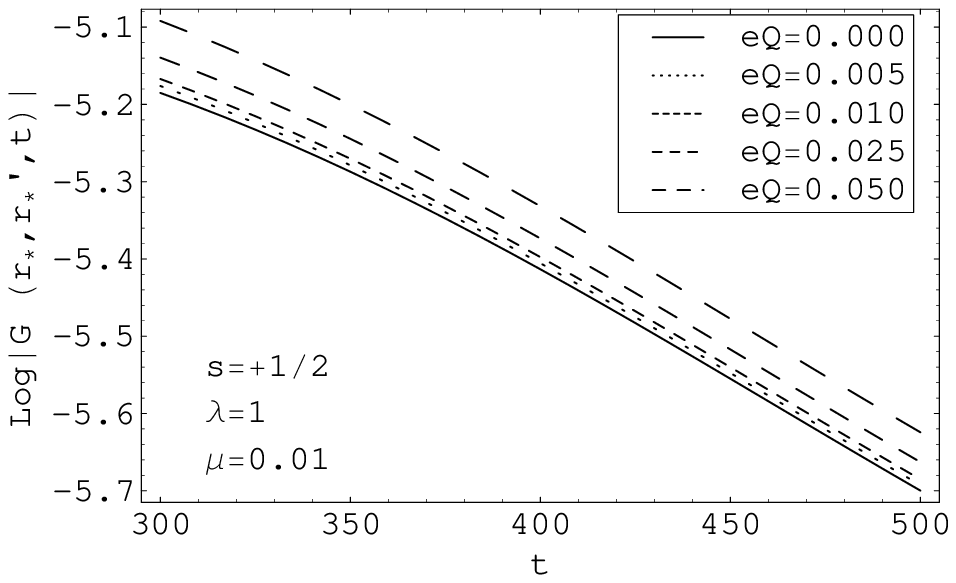}
\includegraphics[scale=0.8]{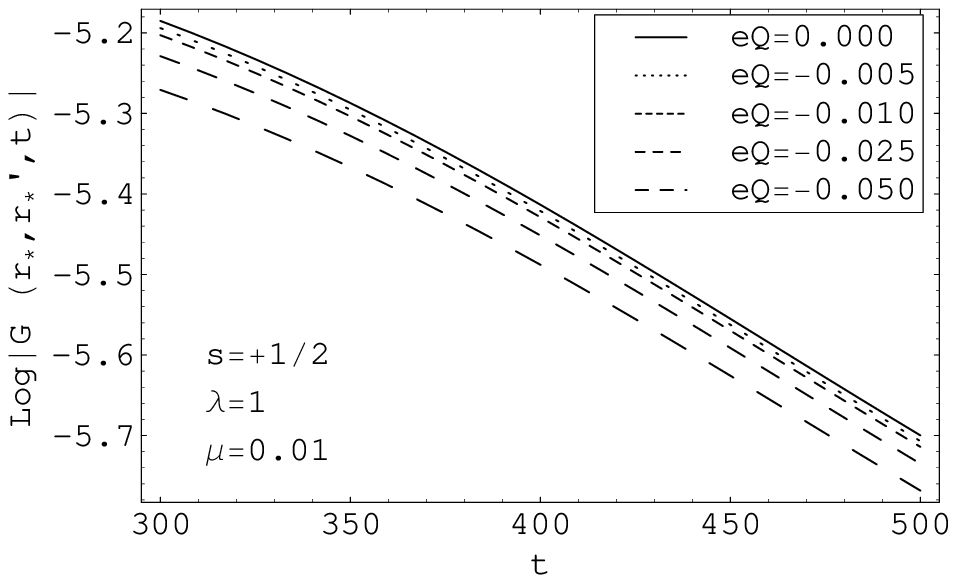} 
\caption{\label{fig1}Graphs of  $\ln|G^C(r_*,r_*';t)|$ versus $t$
for different $seQ$. These figures show that the dumping exponent
depends on the product of the spin weight and the charges of the
black hole and Dirac fields, and $seQ<0$ speeds up the
perturbation decay but $seQ>0$ slows it down.}
\end{figure}

\begin{figure}
\includegraphics[scale=0.8]{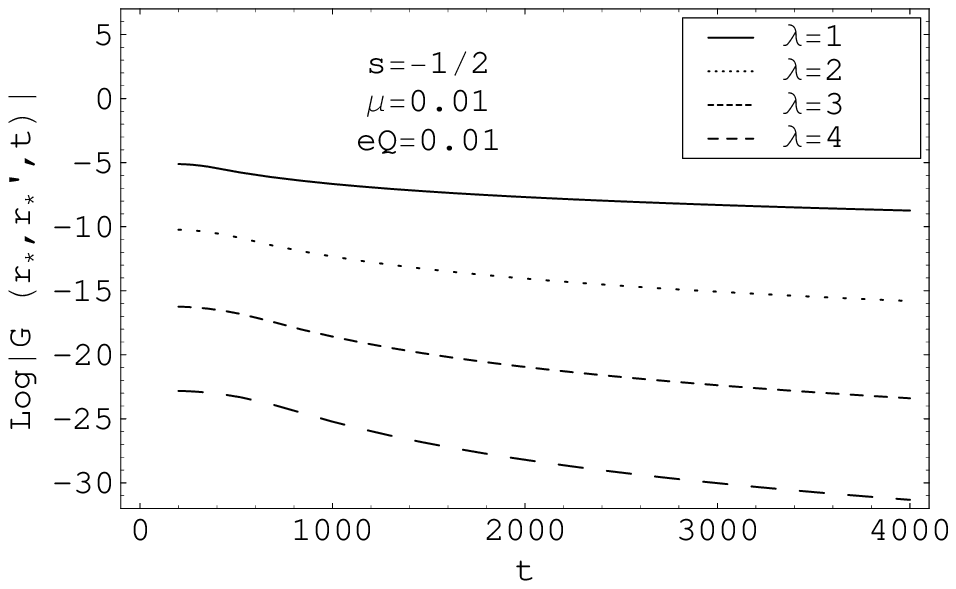}
\includegraphics[scale=0.8]{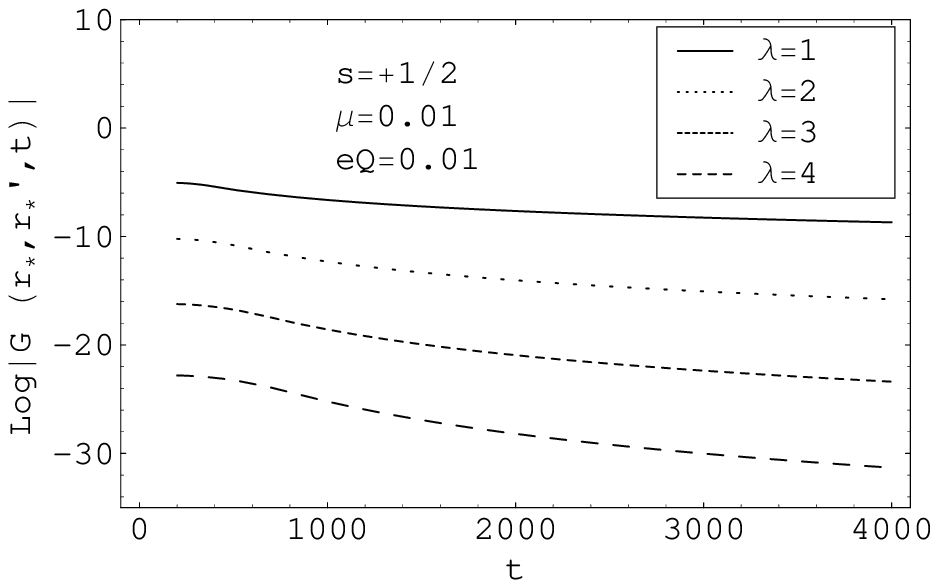}\\
\includegraphics[scale=0.8]{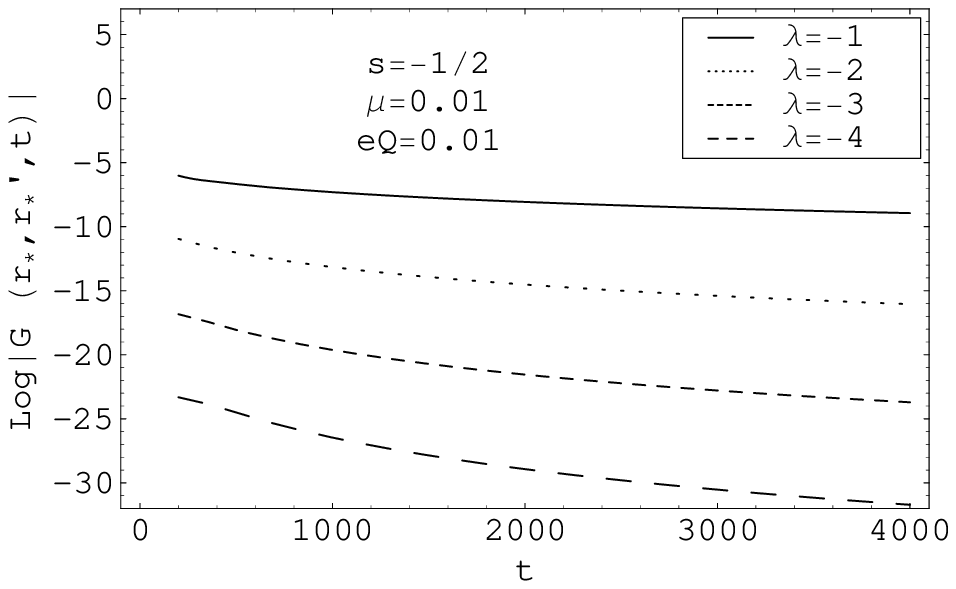}
\includegraphics[scale=0.8]{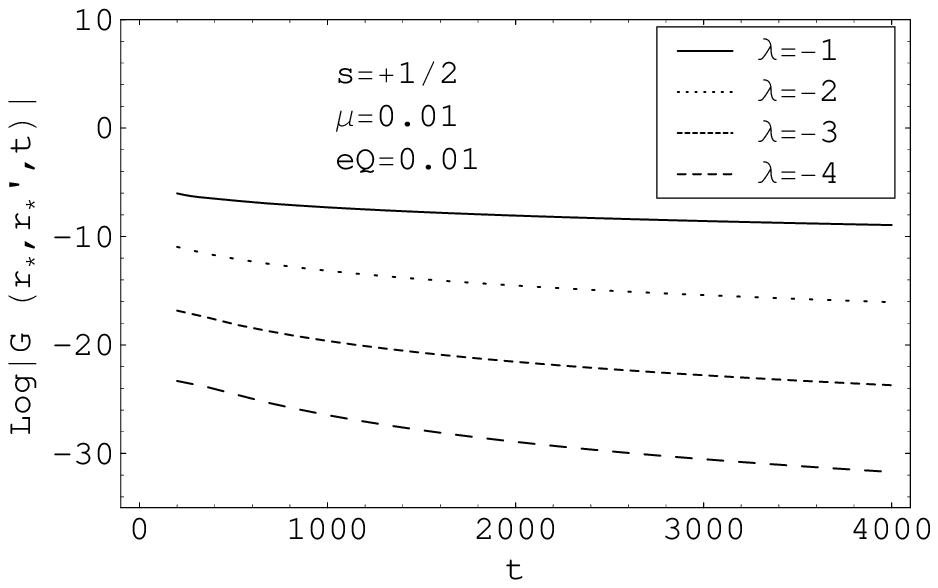}
\caption{\label{fig2}Graphs of $\ln|G^C(r_*,r_*';t)|$ versus $t$
for different $\lambda$ with $s=\pm 1/2$, $\mu=0.01$ and
$eQ=0.01$, showing that the dumping exponent depends on the
multiple number of the wave mode, and the larger the magnitude of
multiple number, the more quickly the perturbation decays.}
\end{figure}

\begin{figure}
\includegraphics[scale=0.8]{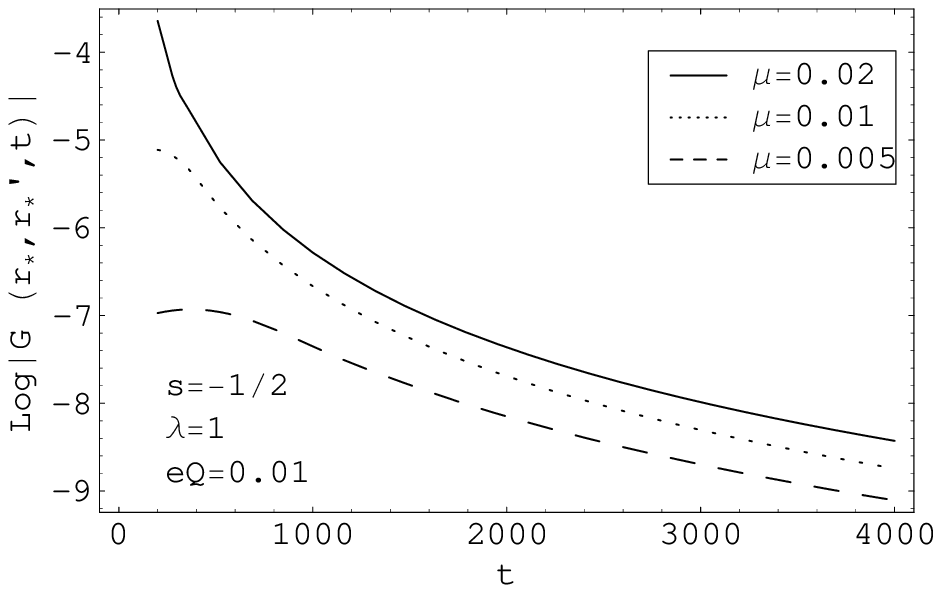}
\includegraphics[scale=0.8]{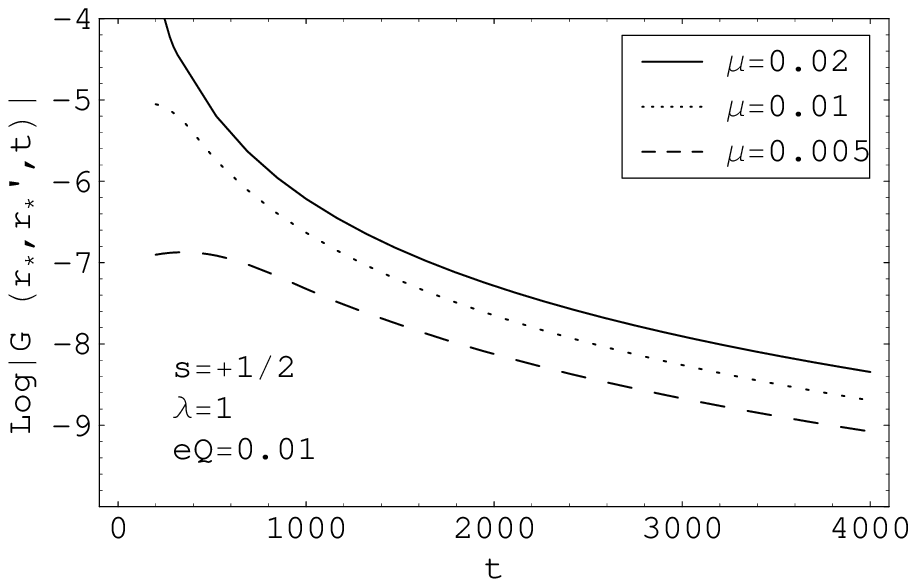} \\
\includegraphics[scale=0.8]{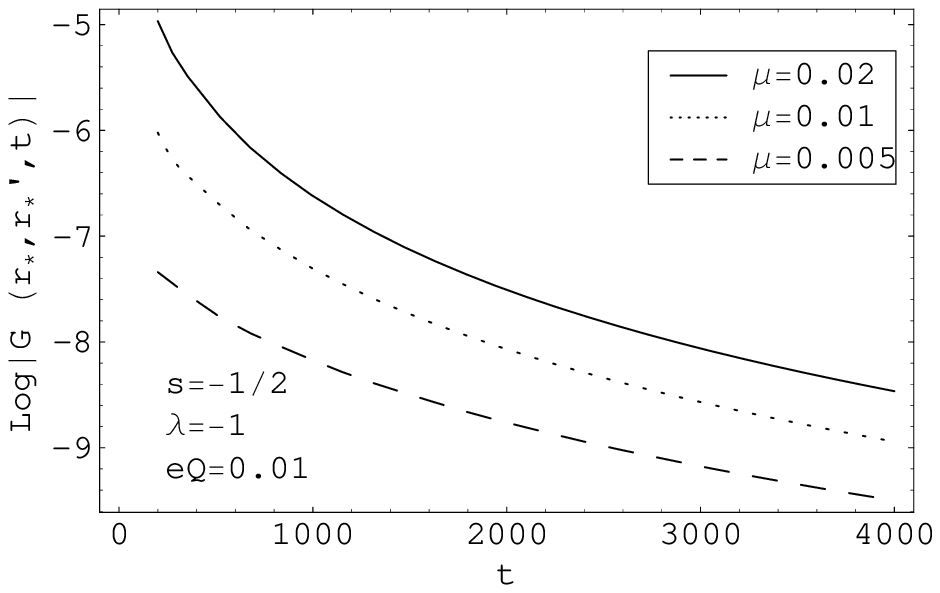}
\includegraphics[scale=0.8]{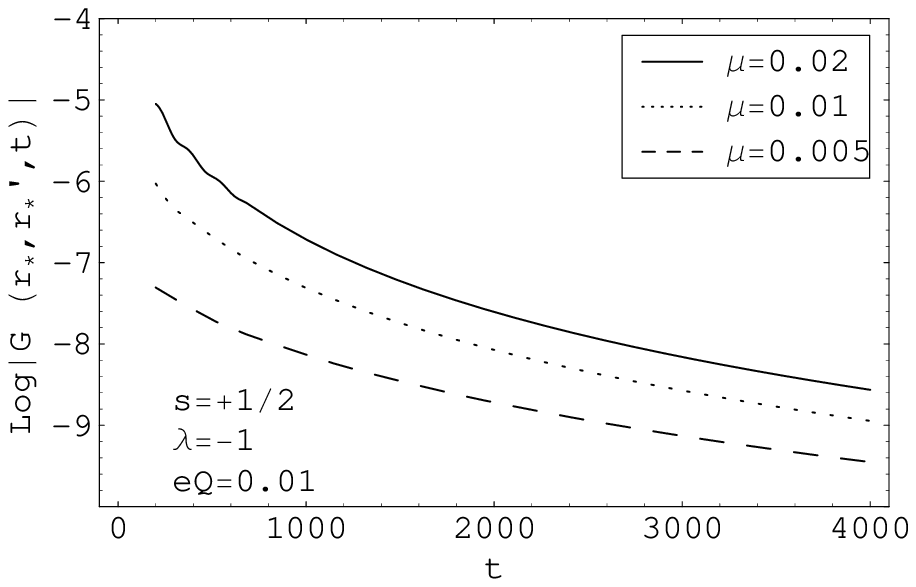}
\caption{\label{fig3}Graphs of $\ln|G^C(r_*,r_*';t)|$ versus $t$
for different mass $\mu$.  The result obtain from these figures is
that the dumping exponent depends on the mass of the Dirac fields,
and the smaller the mass $\mu$, the faster the perturbation
decays.}
\end{figure}

We first focus our attention on the intermediate asymptotic
behavior of the charged massive Dirac fields. That is the tail in
the range $M\ll r\ll t \ll M/(M\mu)^2$. In this time scale, we
find that the frequency range $\varpi=O(\sqrt{\mu/t})$, which
gives the dominant contribution to the integral, implies $a\ll 1$.
Equation (\ref{Pq}) shows that $a$ originates from the $1/r$ term
which describes the effect of backscattering off the spacetime
curvature. That is to say, the backscattering off the spacetime
curvature from the asymptotically far regions is negligible for
the case $a\ll 1$. Then, using the fact that
$M(\tilde{a},\tilde{b},z)\approx 1$ as $z\rightarrow 0$, we have
\begin{eqnarray}
\label{F3}F(\varpi)& =&\frac{\pi}{sin(\pi b)}\frac{1+e^{(1+2b)i
\pi}}{2^{1+2b}b^2}\frac{\varpi^{2b}}{\Gamma(b)^2}(r_*r_*')^{1/2+b}.
\end{eqnarray}
Substituting Eq. (\ref{F3}) into the Eq. (\ref{mgreen1}), we find
\begin{eqnarray}
\label{mgreen2}
 &&G^C(r_*,r_*';t) =\nonumber \\
&&\int_{-\mu}^\mu \frac{(1+e^{(1+2b)i
\pi})(r_*r_*')^{\frac{1}{2}+b}}{ 2^{2b+1}b^2 \Gamma(b)^2 sin(\pi
b)}(\mu^2-\omega^2)^{b}e^{-i\omega t}d \omega.
\end{eqnarray}
Unfortunately, the integral (\ref{mgreen2}) can not be evaluated
analytically since the parameter $b$ depends on $\omega$. However,
we can work out the integral numerically and the corresponding
results are presented in Figs. \ref{fig1}-\ref{fig3}. Figure
\ref{fig1} describes $\ln|G^C(r_*,r_*';t)|$ versus $t$ for
different $eQ$, which shows that the dumping exponent depends on
$seQ$, i.e., the product of the spin weight of the Dirac fields
and the charges of the black hole and Dirac fields, and $seQ<0$
speeds up the decay of the perturbation  but $seQ>0$ slows it
down. Figure \ref{fig2} illustrates $\ln|G^C(r_*,r_*';t)|$ versus
$t$ for different $\lambda$ with $s=\pm 1/2$, $\mu=0.01$ and
$eQ=0.01$, which indicates that the dumping exponent depends on
the multiple number of the wave mode, and the larger the magnitude
of the multiple number, the more quickly the perturbation decays.
Figure \ref{fig3} gives $\ln|G^C(r_*,r_*';t)|$ versus $t$ for
different mass $\mu$ of the Dirac fields, which shows that the
dumping exponent depends on the mass of the Dirac fields, and the
smaller the mass $\mu$, the faster the perturbation decays.

In the above discussion we have used the approximation of ~$a \ll
1$, which only holds when ~$\mu t \ll 1/(\mu M)^{2}$. Therefore,
the power-law tail found earlier is not the final one, and a
change to a different pattern of decay is expected when $a$ is not
negligibly small. Here we examine the asymptotic tail behavior at
very late times such that $\mu t\gg 1/(\mu M)^{2}$. This
asymptotic tail behavior is caused by a resonance backscattering
due to spacetime curvature \cite{14}. In this case, we have $a
\simeq (M\mu^2+ e Q\mu)/\varpi \gg 1$, namely, the backscattering
off the spacetime curvature in asymptotically far regions is
important. Using Eq. (13.5.13) of Ref. \cite{Abramowitz} and Eq.
(\ref{FF}), we obtain
\begin{eqnarray}\label{FFF}
&&F(\omega)\simeq\frac{\Gamma(1+2 b)\Gamma(1-2b)r_*' r_*}{2
b}\nonumber \\ &&[J_{2b}(\sqrt{\alpha r_*'})J_{-2b}(\sqrt{\alpha
r_*}) -I_{2b}(\sqrt{\alpha r_*'})I_{-2b}(\sqrt{\alpha
r_*})]\nonumber \\
&&+\frac{\Gamma(1+2 b)^2\Gamma(-2b)r_*' r_*}{2 b \Gamma(2b)}
 \frac{\Gamma(\frac{1}{2}+b-a)}{2 b \Gamma(\frac{1}{2}-b-a)}
a^{-2b} \nonumber \\
&&[J_{2b}(\sqrt{\alpha
r_*'})J_{2b}(\sqrt{\alpha r_*})+I_{2b}(\sqrt{\alpha
r_*'})I_{2b}(\sqrt{\alpha r_*})],
\end{eqnarray}
where $\alpha=8(M\mu^2+e Q \mu)$, and $I_{\pm(2b)}$ is the
modified Bessel function.

\begin{figure}
\includegraphics[scale=1.8]{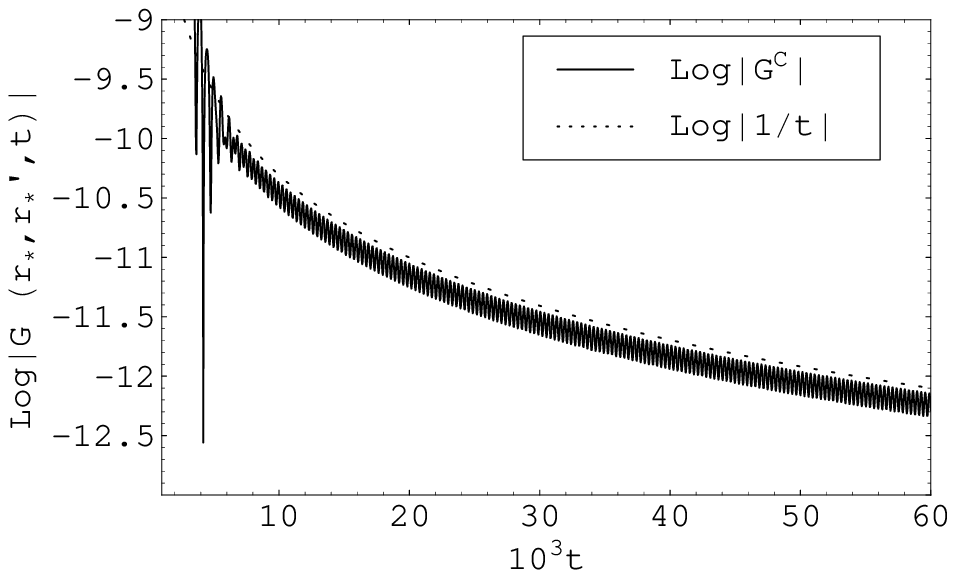}
\caption{\label{fig4}Graphs of $\ln|G^C(r_*,r_*';t)|$ versus $t$
with $M=100$, $\mu=0.01$, $s=-1/2$, $\lambda=1$, $Q=0.1M$ and
$e=0.002$  for the first term (solid line). The dashed line is
$\sim \log \left|1/t\right|$.  }
\end{figure}

We can study late-time behavior for the first term of Eq.
(\ref{FFF}) using numerical method.  The result is presented by
Fig. \ref{fig4} which shows that asymptotically late-time tail
arising from the first term is still $\sim t^{-1}$ although the
factor $\Gamma(1+2 b)\Gamma(1-2b)/2b$ is not a constant.

Now let us to find the behavior of the second term. Because of
$\Gamma(1+2 b)\Gamma(1-2b)/2b= -\Gamma(1+2
b)^2\Gamma(-2b)/(2b\Gamma(2b)) $ and the factor $\Gamma(1+2
b)\Gamma(1-2b)/2b$ almost dose not affect the asymptotical
behavior of the first term, we can define
\begin{eqnarray}
C&=&\frac{\Gamma(1+2 b)^2\Gamma(-2b)r_*' r_*}{2 b
\Gamma(2b)}[J_{2b}(\sqrt{\alpha r_*'})J_{2b}(\sqrt{\alpha
r_*})\nonumber \\ && +I_{2b}(\sqrt{\alpha
r_*'})I_{2b}(\sqrt{\alpha r_*}),\nonumber
\end{eqnarray}
which approximates to a constant. Then, in the limit $a\gg 1$, the
contribution of the second term to the Green's function can be
expressed as
\begin{eqnarray}
G^C(r_*,r_*';t)\sim \frac{C}{2\pi}\int_{-\mu}^{\mu} e^{i(2\pi
a-\omega t)}e^{i\phi}  d\omega, \label{Gasym1}
\end{eqnarray}
where the phase $\phi$ is determined by $
e^{i\phi}=\frac{1+(-1)^{2b}e^{-2i\pi a}}{1+(-1)^{2b}e^{2i\pi a}},$
and it remains in the range $0\leq \phi \leq 2\pi$, even if $a$
becomes very large.  We use saddle-point integration to evaluate
Eq. (\ref{Gasym1}) and find
\begin{eqnarray}
&&G^C(r_*,r_*';t)\sim
\frac{C\mu}{2\sqrt{3}}\left(2\pi\right)^{\frac{5}{6}}(M\mu+2e
Q)^{\frac{1}{3}} (\mu t)^{-\frac{5}{6}}\nonumber \\
&&  sin \{ \mu t -[2\pi(M\mu+2 e Q)]^{\frac{2}{3}}(\mu
t)^{\frac{1}{3}}-\phi(\omega_0)-\frac{\pi}{4}\}, \label{Gasym2}
\end{eqnarray}
which is the asymptotic behavior of the Green's function at very
late times. Eq. (\ref{Gasym2}) shows that the decay rate of the
asymptotic tail is $t^{-5/6}$ and the oscillation of the tail has
the period $2\pi/\mu$ which is modulated by two types of long-term
phase shifts, a monotonously increasing phase shift $[2\pi(M\mu+2
e Q)]^{\frac{2}{3}}(\mu t)^{\frac{1}{3}}$ and a period phase shift
$\phi(\omega_0)$.

\begin{figure}
\includegraphics[scale=1.8]{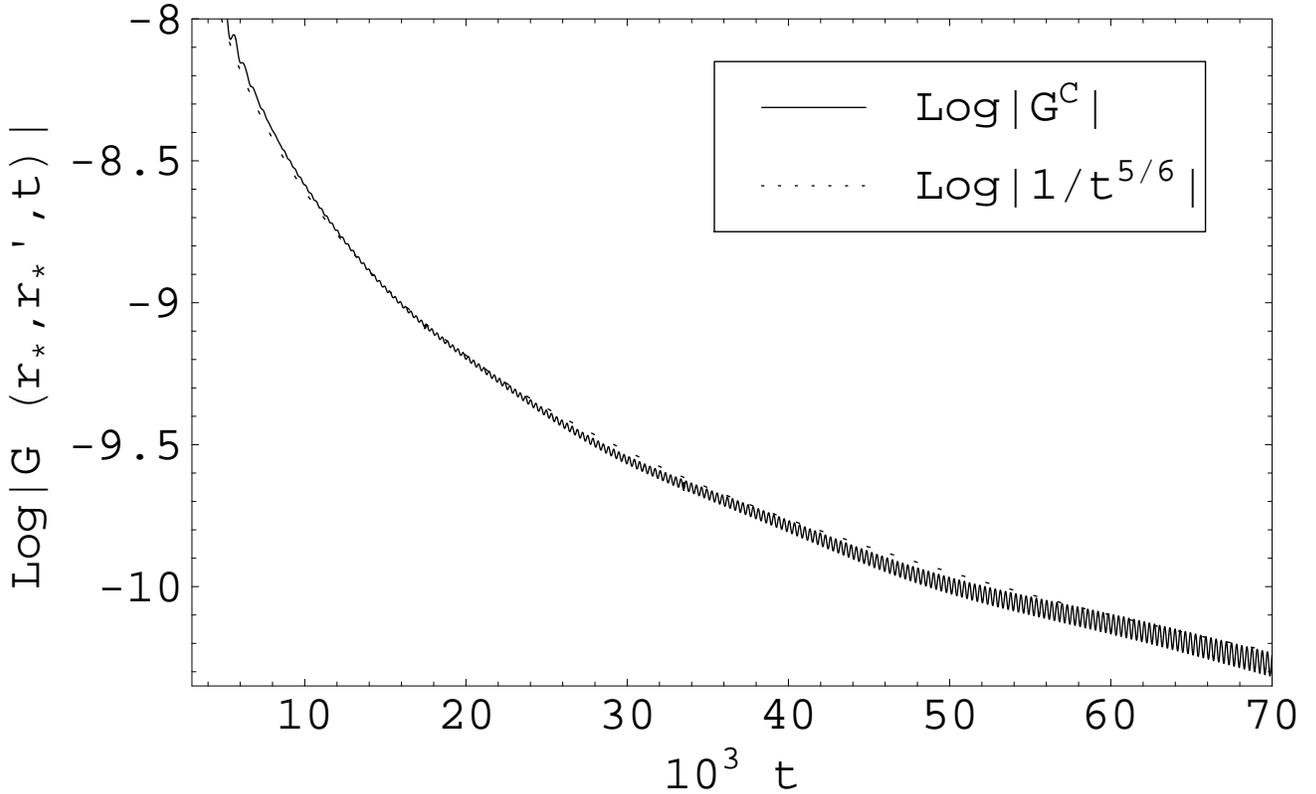}
\caption{\label{fig5}Graphs of $\ln|G^C(r_*,r_*';t)|$ versus $t$
with $M=100$, $\mu=0.01$, $s=-1/2$, $\lambda=1$, $Q=0.1M$ and
$e=0.002$  for the second term (solid line). The dashed line is
$\sim \log \left|t^{-5/6}\right|$.  }
\end{figure}

To confirm the analytical prediction, we present the numerical
result of the second term in Fig. \ref{fig5} and find that the
decay rate of the asymptotic tail is still $t^{-5/6}$.

\section{summary}

The intermediate late-time tail and the asymptotic tail behavior
of the charged massive Dirac fields in the background of the RN
black hole are studied. The results of the intermediate late-time
tail are presented by figures because we can not obtain
analytically Green's function $G^C(r_*, r_*'; t)$ because the
parameter $b$ in the integrand of the Green's function depends on
the integral variable $\omega$. We learn from the figures that the
intermediate late-time behavior is dominated by an inverse
power-law decaying tail without any oscillation, which is
different from the oscillatory decaying tails of the scalar
fields. It is interesting to note that the dumping exponent
depends not only on the multiple number of the wave mode but also
on the mass of the Dirac fields and the product $seQ$.  We also
find that the decay rate of the asymptotically late-time tail is
$t^{-5/6}$ and the oscillation of the tail has the period
$2\pi/\mu$ which is modulated by two types of long-term phase
shifts.

\begin{acknowledgments}
This work was supported by the National Natural Science Foundation
of China under Grant No. 10473004; the FANEDD under Grant No.
200317; the SRFDP under Grant No. 20040542003; and the Hunan
Provincial Natural Science Foundation of China under Grant No.
04JJ3019.
\end{acknowledgments}

\end{document}